\RequirePackage{fix-cm}
\documentclass[smallextended]{svjour3}       
\smartqed  
\usepackage{graphicx}
\usepackage{amsmath,amsfonts,amssymb}
\usepackage{color}

\newcommand{\mb}[1]{\mathbf #1}
\newcommand{\mc}[1]{\mathcal #1}

\newcommand{\eq}[1]{Eq.~\eqref{eq:#1}}
\newcommand{\fig}[1]{Fig.~\ref{fig:#1}}
\newcommand{\sect}[1]{Sec.~\ref{sec:#1}}

\begin{document}

\title{Fermi surfaces of composite fermions}


\author{R. N. Bhatt         \and
        Matteo Ippoliti 
}


\institute{R. N. Bhatt \at
              Department of Electrical Engineering, Princeton University, Princeton, NJ 08544 \\
              \email{ravin@princeton.edu}           
           \and
           Matteo Ippoliti \at
           Department of Physics, Princeton University, Princeton, NJ 08544 
}

\date{Received: July 13, 2019 / Accepted: date}

\maketitle

\begin{abstract}
The fractional quantum Hall (FQH) effect was discovered in two-dimensional electron systems subject to a large perpendicular magnetic field nearly four decades ago.  
It helped launch the field of topological phases, and in addition, because of the quenching of the kinetic energy, gave new meaning to the phrase ``correlated matter''. 
Most FQH phases are gapped like insulators and superconductors; however, a small subset with even denominator fractional fillings $\nu$ of the Landau level, typified by $\nu=1/2$, are found to be gapless, with a Fermi surface akin to metals. 
We discuss our results, obtained numerically using the infinite-Density Matrix Renormalization Group (iDMRG) scheme, on the effect of non-isotropic distortions with discrete $N$-fold rotational symmetry of the Fermi surface at zero magnetic field on the Fermi surface of the correlated $\nu = 1/2$ state. 
We find that while the response for $N = 2$ (elliptical) distortions is significant (and in agreement with experimental observations with no adjustable parameters), it decreases very rapidly as $N$ is increased. 
Other anomalies, like resilience to breaking the Fermi surface into disjoint pieces, are also found.
This highlights the difference between Fermi surfaces formed from the kinetic energy, and those formed of purely potential energy terms in the Hamiltonian.
\end{abstract}

\section{Introduction}

Following the experimental discovery~\cite{Tsui1982} of the FQH effect at an electron density corresponding to a filling of $\nu = 1/3$ in the lowest Landau level, Laughlin~\cite{Laughlin1983} provided an understanding of the existence of FQH liquid phases at other inverse odd-integer fillings, e.g. 1/5, 1/7 etc. 
Theoretical work~\cite{Haldane1983}~\cite{Jain1989}~\cite{Jain2007} categorized the large number of FQH phases seen in subsequent experiments~\cite{Du1993}~\cite{Stormer1999} into a hierarchy of fractions. 
The most widely used scheme today~\cite{Jain2007} relies on the concept of ``composite fermions'' (CFs), i.e. electrons bound to an even number of flux quanta or vortices. 
A mean-field description of the electron system at $\nu = 1/2$ in terms of composite fermions made up of electrons bound to two flux quanta gives rise to a Fermi liquid-like state~\cite{Halperin1993} known as the composite Fermi liquid (CFL). 
This result has been verified in a number of experiments including direct determination of the Fermi surface~\cite{Kamburov2012B}. 
Nevertheless, it is is a very counterintuitive result in terms of the original electron system for which the kinetic energy is totally quenched by Landau quantization, and the Hamiltonian consists of purely potential terms! 
For systems possessing continuous rotational symmetry in two dimensions, the composite Fermi surface is a circle, as might be expected, with a radius determined by  Luttinger's theorem.

It has been recognized in recent years~\cite{Son2015} that this composite Fermi surface is characterized by an unusual Berry phase, which may point to the composite fermions being massless Dirac particles, and was subsequently verified numerically with various methods~\cite{Geraedts2016}~\cite{Geraedts2018}. 
Here, we consider a different, but equally interesting property of the composite fermion Fermi surface, namely its response to terms in the Hamiltonian that break rotational symmetry. This explores the geometric degree of freedom of quantum Hall states~\cite{Haldane2011}.

\begin{figure}
\centering
\includegraphics[width=\textwidth]{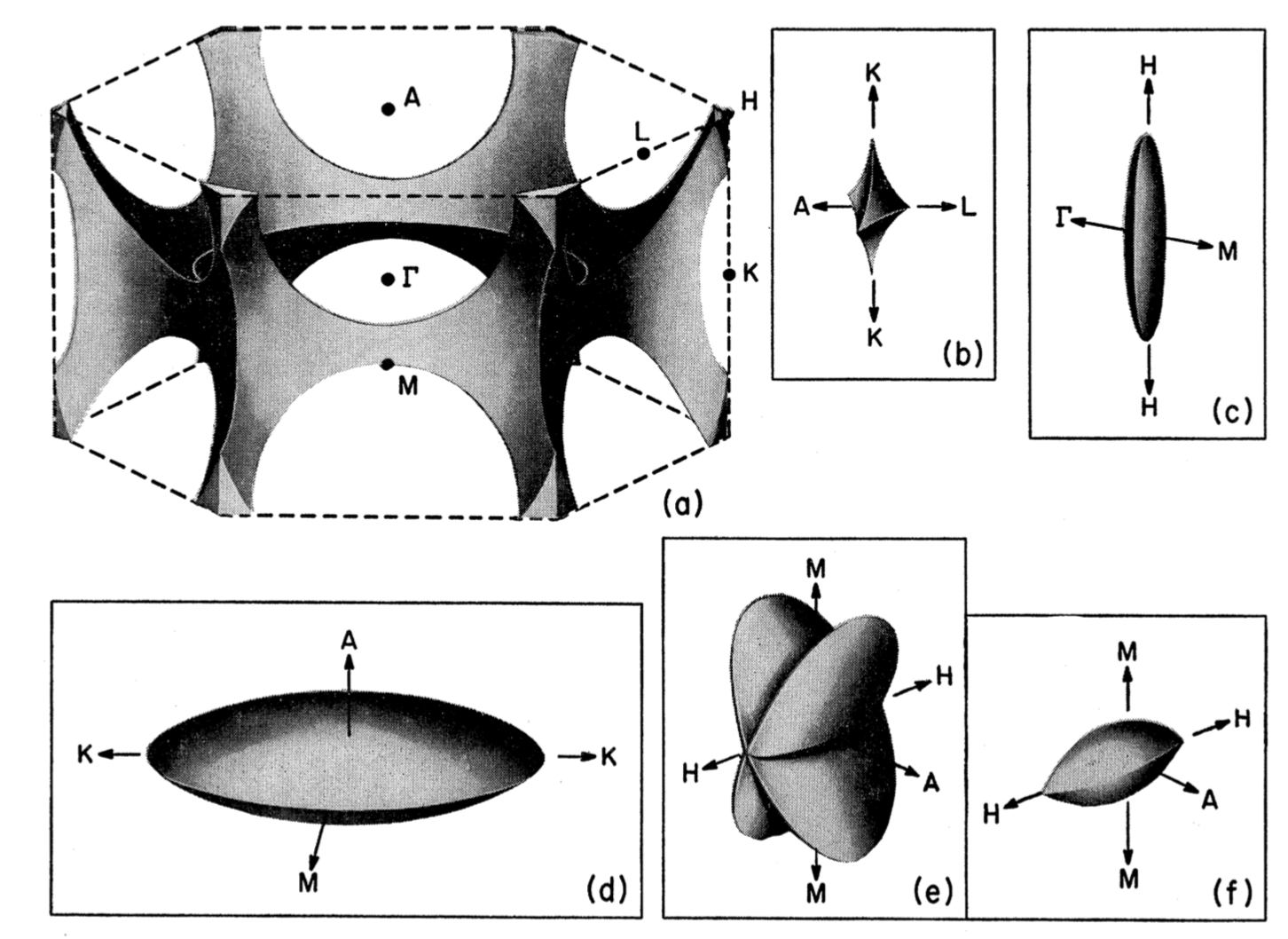}
\caption{Fermi surfaces of solid Magnesium. From Ref.~\cite{Ketterson1967}.\label{fig:sf} }
\end{figure}

Energy bands in solids depend significantly on the atomic structure of the constituents through the lattice symmetry and the effective pseudopotentials; consequently, Fermi surfaces of metals show tremendous variety. 
Deducing them using various experimental probes was a major area of study in the 1960s, which proved crucial to determine the validity of different theoretical approximations for calculating energy bands. 
While the alkali metals have a nearly spherical surface with small distortions, and the noble metals have spherical Fermi surfaces with multiple necks, others can be quite complex. 
\fig{sf}, taken from Ref.~\cite{Ketterson1967}, depicts the Fermi surface of solid Magnesium, with colorful descriptions of various parts -- monster, cap, cigar, butterfly {\it etc.}.
For another complex Fermi surface, see Ref.~\cite{Ashcroft1976}.

\begin{figure}
\centering
\includegraphics[width=0.8\textwidth]{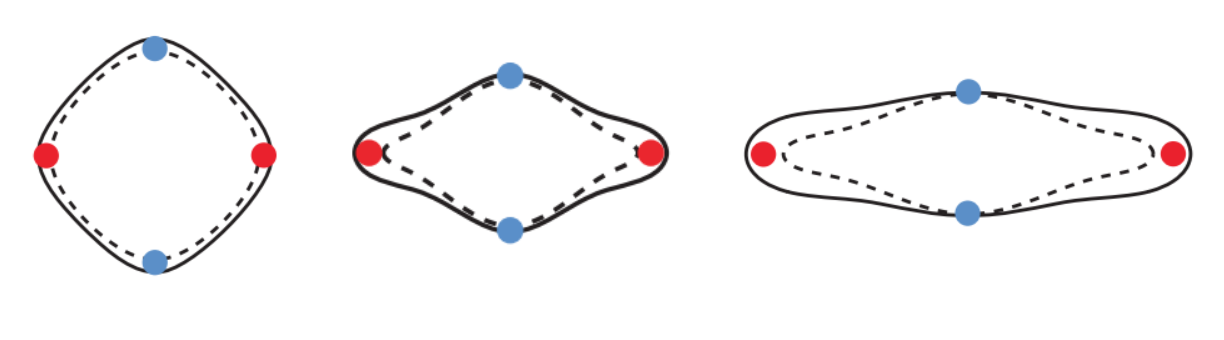}
\caption{Fermi surfaces of holes in strained GaAs. 
Solid and dashed lines represent calculated values for different spin components. 
Dots represent results obtained from experimentally measured oscillations in the resistivity as a result of commensuration effects.
From Ref.~\cite{Jo2017}.\label{fig:fs2} }
\end{figure}

Given the extreme variability of Fermi surfaces of various metallic materials to atomic potentials and lattice structure, it is natural to ask how the circular Fermi surface of the correlated $\nu = 1/2$ state for isotropic systems responds to terms in the Hamiltonian that do not preserve rotational symmetry. 
In particular, we investigated the effect of having a discrete $N$-fold symmetry in the zero-field kinetic energy, which could arise as a result of lattice symmetry, on the shape of the Fermi surface at high magnetic field at $\nu = 1/2$. 
Surprisingly, while we found the Fermi surface distorts easily from a circular to an elliptical shape (in a manner that was quantitatively in agreement with experiments on strained GaAs quantum wells~\cite{Jo2017}, see Fig.~\ref{fig:fs2}), other types of distortions elicited a much more subtle response.
Thus, as we show analytically, the composite fermion Fermi surface is completely insensitive to isotropic distortions, even when these may split the zero-field Fermi sea into disconnected ring-shaped pieces.
In an anisotropic situation, even when the zero-field Fermi surface splits into well-separated Fermi pockets, we found that the composite Fermi surface remains a single connected piece, a result also found in experiment.~\cite{Kamburov2014} 
Finally, we found the composite Fermi surface to be extremely resilient to distortions with four-fold and higher rotational symmetry.

\section{Mapping the Fermi contour of composite fermions \label{sec:method}}

We now describe how to map the Fermi contour of composite fermions numerically with the infinite density matrix renormalization group (iDMRG) method~\cite{Schollwock2011}.
This method, originally developed for one-dimensional spin chains, has been successfully applied to the fractional quantum Hall problem~\cite{Zaletel2013}~\cite{Zaletel2015}.
By placing the system on a cylinder, Landau orbitals wrapping around the finite circumference $L$ can be mapped onto sites of an infinite spin chain. 
This geometry is crucial to the method we are about to outline.
Beyond the finite circumference $L$, DMRG naturally introduces a second cutoff, known as the \emph{bond dimension} $\chi$. 
This is the size of the matrices used in the variational \emph{ansatz} for the many body ground state, and can be thought of as a limit on the amount of quantum entanglement in the state. 
For gapped ground states (whose entanglement obeys an area law, i.e. it stays finite even on the infinite cylinder), the error induced by this approximation drops quickly with $\chi$~\cite{Verstraete2006}, so in principle there exists a value $\chi^\ast$ above which convergence within any predefined error bar is achieved.
This is not true of gapless ground states, where the entanglement diverges logarithmically. 
As a result, this method may appear ill-suited to the study of the CFL.
However, as we shall see shortly, that is not the case.

The method we use, pioneered in Ref.~\cite{Geraedts2016}, relies on the static guiding center structure factor, 
\begin{equation}
S(\mb q) \equiv \langle \bar{\rho}_{\mb q} \bar{\rho}_{-\mb q} \rangle \; ,
\end{equation}
where $\bar{\rho}_{\mb q} \equiv \sum_j e^{i\mb q\cdot \mb R_j}$ is the LLL-projected density operator, expressed in terms of the guiding center operators $\mb R_j$ ($j$ labels the electrons in the system).
This quantity plays a crucial role in various aspects of quantum Hall physics, e.g., the incompressibilty of FQH fluids is manifested in the long-wavelength scaling $S(q) \sim q^4$~\cite{Girvin1985}. 
In the CFL state, which is compressible, $S(\mb q)$ is not expected to exhibit such quartic behavior.
On the other hand, it displays a singular feature at $q=2k_F$, where $k_F$ the Fermi wavevector of composite fermions.
These singularities allow us to map the Fermi contour of CFs.

\begin{figure}
\centering
\includegraphics[width=\textwidth]{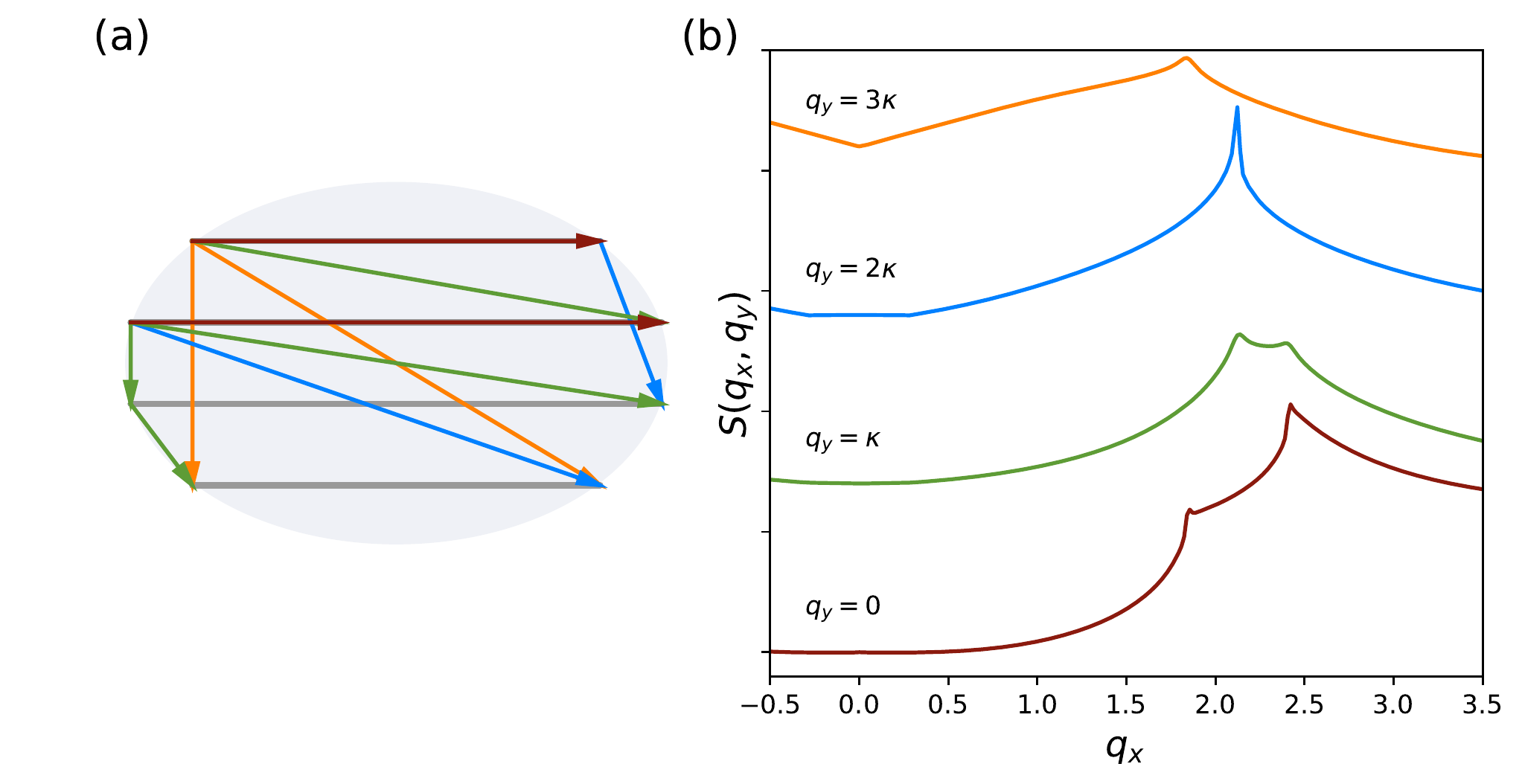}
\caption{
Mapping the CFL Fermi contour from the guiding center structure factor $S(\mb q)$.
(a) Fermi sea for the composite fermions at $L = 17 \ell_B$ and $\alpha_F = 0.445$. 
The shaded area represents the 2D Fermi sea in the planar ($L \to \infty$) limit.
The gray lines contain allowed values of $\mb q$ on the cylinder with finite $L$.
Colored arrows show momenta connecting points on the Fermi contour.
(b) Numerical data for $S(\mb q)$ at $q_y = n\kappa$, $0\leq n \leq3$, with bond dimension $\chi = 3000$, translated vertically for clarity.
The location of the singularities in (b) can be used to reconstruct the shape in (a), initially not known. The CF anisotropy in this case is found to be $\alpha_{CF} = 0.667$.
\label{fig:method} }
\end{figure}

By taking the $x$ axis parallel to the cylinder axis and $y$ around the finite circumference, the component $q_y$ is discretized in steps of $\kappa \equiv 2\pi/L$, while $q_x$ can in principle take on continuous values.
Therefore at any finite $L$ the CF Fermi sea consists of a sequence of one-dimensional segments of lengths $\{ Q_n \}$. 
The Fermi contour then consists of the isolated endpoints of such segments,
$\{\mb q = \left(\pm Q_n/2, \kappa (n+\delta)\right)\}$.~\footnote{$\delta$ may be 0 or $1/2$ depending on the boundary conditions (periodic or antiperiodic) of CFs, which are emergent.}
$S(\mb q)$ is expected to show a non-analytical feature whenever $\mb q$ connects two different points belonging to the Fermi contour.
Locating these singularities allows us to extract the $Q_n$ momenta and effectively map the Fermi contour by ``connecting the dots''.
An example of this procedure is shown in \fig{method}.
The ellipse in \fig{method}(a) represents the anisotropic Fermi sea whose shape we want to determine. 
The colored arrows represent vectors $\Delta\mb q$ connecting points on the Fermi contour.
If we fix $\Delta q_y=0$ (horizontal arrows), we expect singularities in $S(\Delta q_x,0)$ at $\Delta q_x = Q_m$. 
Such singularities are indeed visible in the data of \fig{method}(b), and are in principle all we need to extract the $\{ Q_n \} $ values and map the Fermi contour.
However, as a consistency check, we also find singularities at other values of $\delta q_y$, corresponding to vectors with a non-zero vertical component:
$\Delta \mb q = \left( (Q_m \pm Q_n)/2 , \kappa(m-n) \right)$, for $m\neq n$.
A given estimate of the singularities $\{Q_m\}$ at $\Delta q_y = 0$ naturally comes with a prediction for the whole set of singularities at $|\Delta q_y| > 0$, 
which provides a robust test of the existence of the Fermi contour.

Since the states obtained by DMRG are approximations of the true ground state, these singularities will not be reproduced perfectly. 
In particular, the finite bond dimension $\chi$ introduces a finite correlation length $\xi$ (whereas the ``true'' CFL ground state should have algebraic correlations). 
As a result, the singular features in $S(\mb q)$ acquire finite width $\delta q \sim \xi^{-1}$. 
In practice, for the sizes we study, $\delta q$ can be made small enough with $\chi \leq 6000$ that the gapless nature of the CFL is not the dominant source of uncertainty of the method.

\section{Band mass anisotropy \label{sec:bandmass}}

With the method outlined above, we can test the effects of an anisotropic band mass in the kinetic energy of the zero-field carriers,
\begin{equation}
H_0 = \frac{p_x^2}{2m_{xx}}  + \frac{p_y^2}{2 m_{yy}} 
\equiv \frac{1}{2m} \left( \alpha_F p_x^2 + \alpha_F^{-1} p_y^2 \right) \;,
\end{equation}
where $m = \sqrt{m_{xx} m_{yy}} $ and $\alpha_F$ is the anisotropy parameter, equal to the ratio of extremal Fermi wavevectors, i.e. the square root of mass anisotropy, 
\begin{equation}
\alpha_F = \frac{k_{F,y}}{k_{F,x}} = \sqrt{\frac{m_{yy}}{m_{xx}}} \;.
\end{equation}

\begin{figure}
\centering
\includegraphics[width=0.65\textwidth]{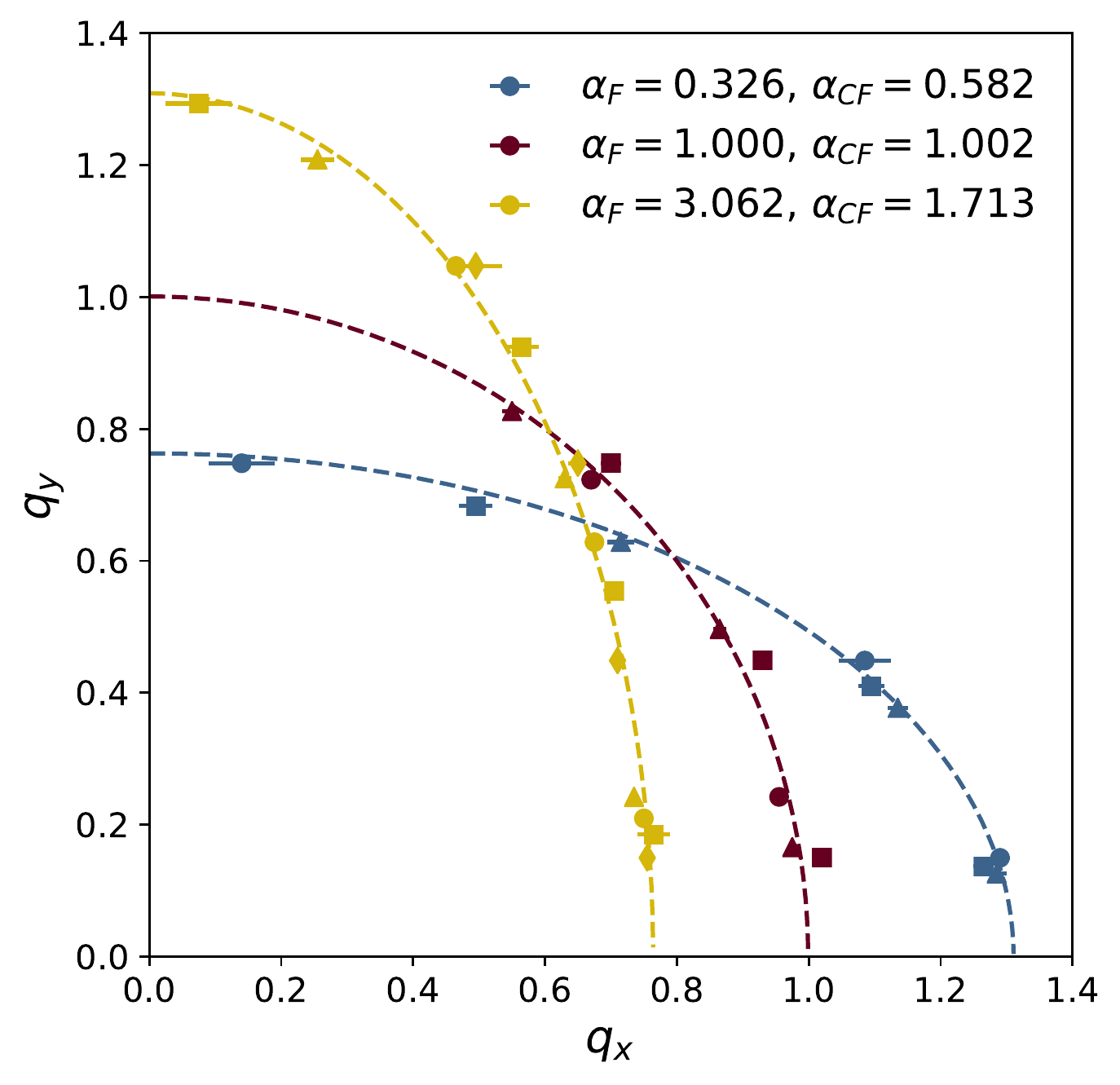}
\caption{
Location of the Fermi contour for three selected values of $\alpha_F$, extracted from data similar to that in \fig{method}. 
The dashed lines are the best fits of the data to ellipses of fixed area $\pi \ell_B^{-2}$, with the value of $\alpha_{CF}$ shown in the legend. 
Different symbols of the same color (circle, square, triangle and diamond) correspond to different system sizes for the same $\alpha_F$.}
\label{fig:ellipses}
\end{figure}

We consider values of $\alpha_F$ ranging from $0.16$ to $6.25$. 
At each value of $\alpha_F$, we map the Fermi contour as described earlier, using several values of $L$ (3 to 5 distinct values in the range of $13$ to $27 \ell_B$, depending on $\alpha_F$).
Examples representative of the cases $\alpha_F <1$, $\alpha_F = 1$ and $\alpha_F >1$ are shown in \fig{ellipses}.
We extract $\alpha_{CF}$ from the optimal fit of the Fermi contour points to an ellipse. 
Though in the thermodynamic limit we expect an elliptical Fermi contour, our finite-size data necessarily deviate from it due to Luttinger's theorem, which in our system fixes the total length of segments making up the finite-size Fermi sea to 
\begin{equation}
\sum_n Q_n = \nu L \ell_B^{-2} \;.
\label{eq:luttinger}
\end{equation}
This leads to an error in the anisotropy of our fitted ellipse because of the finite values of $L$, which we estimate to be of order $1-2\%$.
We fit the discrete set of $(q_x,q_y)$ coordinates obtained through the process described above to an ellipse of area $\pi \ell_B^{-2}$ (which is fixed by the carrier density at half filling). 
From the optimal ellipse $(k/k_{CF,x})^2 + (k/k_{CF,y})^2 = 1$, 
we obtain $\alpha_{CF} = k_{CF,y}/k_{CF,x}$.

\begin{figure}
\centering
\includegraphics[width=0.7\textwidth]{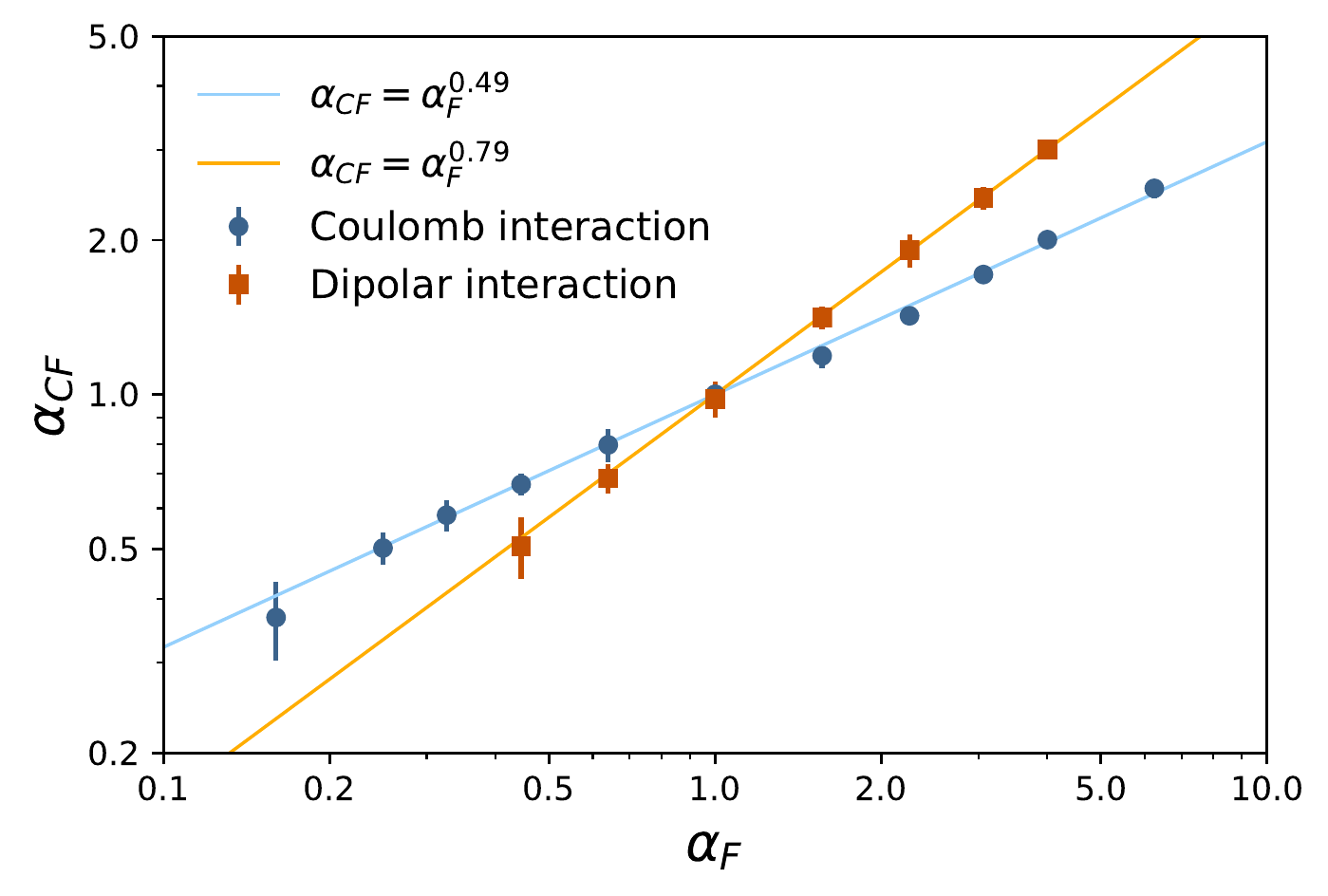}
\caption{
Composite fermion anisotropy $\alpha_{CF}$, obtained by mapping the Fermi contour of the CFL state obtained numerically with iDMRG, as a function of bare electron anisotropy $\alpha_F$ for Coulomb interaction $V(r) = 1/r$ and dipolar interaction $V(r) = 1/r^3$.
The solid lines correspond to power-law fits $\alpha_{CF} = (\alpha_F)^{c_1}$ as shown in the legend.
}
\label{fig:results}
\end{figure}

In the planar limit $L\to\infty$, a $\pi/2$ rotation exchanges the major and minor axes of the elliptical Fermi surface, mapping $\alpha$ to $ 1/\alpha$.
This implies that $\log(\alpha_{CF})$ is an odd function of $\log(\alpha_F)$, 
so its Taylor expansion around the isotropic point $\alpha_F = \alpha_{CF} = 1$ contains only odd powers:
\begin{equation}
\log (\alpha_{CF}) = \sum_{n=0}^{\infty} c_{2n+1}( \log \alpha_F )^{2n+1} = c_1 \log \alpha_F + \mathcal O((\log\alpha_F)^3) \;.
\label{eq:logeqn}
\end{equation}
We find~\cite{Ippoliti2017A} that the first term in this expansion, corresponding to a power law $\alpha_{CF} = (\alpha_F)^{c_1}$, already provides a very good approximation to the data for Coulomb interaction\footnote{This is partly explained by the absence of a quadratic term ($c_2=0$ by symmetry), but also due to the (unexplained, as far as we know) smallness of the symmetry-allowed cubic term $c_3$.}, as shown in \fig{results}.
A linear fit yields $c_1 \simeq 0.49 \pm 0.1$, close to a square-root dependence $\alpha_{CF} = \sqrt{\alpha_F}$.
Remarkably, this is in very good agreement with experiments on holes in GaAs quantum wells under application of in-plane strain~\cite{Jo2017}. 
 
In order to test the universality of this result, we replace the Coulomb interaction with a dipolar interaction $V(r) \propto r^{-3}$, also shown in \fig{results}.
This interaction could be realized in systems of cold atoms.~\cite{Wilkin2000}~\cite{Yao2013}~\cite{Cooper2013}
We find again that a single power law (i.e. a truncation of \eq{logeqn} to the first term) fits the data well, but we measure a significantly larger exponent $c_1 = 0.79 \pm 0.01 $.
This is unambiguously different from a square root dependence.
In particular it implies that $\alpha_{CF}$ is much closer to $\alpha_F$ (i.e. the CFL is much more anisotropic) than for the case of Coulomb interaction.
Since the CFL anisotropy arises from the competition between the kinetic energy term (which has anisotropy $\alpha_F$) and the interaction (which is isotropic),
this result means that the dipolar interaction is \emph{less} effective than the Coulomb one at countering the effect of band mass anisotropy.
This is intuitively reasonable: as $r^{-3}$ decays more quickly than $r^{-1}$, its effect on the shape of long-distance correlations will be weaker. 

\section{Isotropic distortions}
\label{sec:rings}

We now take a brief departure from anisotropic deformations to consider the effect of circularly symmetric, but otherwise arbitrary, deformations of the electron dispersion~\cite{Ippoliti2017B}.
We note in passing that non-parabolic dispersions could occur in any strongly interacting fermionic system with continuous translational and full rotational symmetry ({\it i.e.} Galilean invariance). 
In fact, significant deviations have been predicted and seen experimentally in Galilean invariant three-dimensional dilute mixtures of ${}^3$He in ${}^4$He~\cite{Varma1973}~\cite{Bhatt1978} which depend on the effective interaction pseudopotentials.~\cite{Hsu1985}

We show that the problem in the high field limit reduces to the usual Landau
problem with the \emph{same} extensively degenerate Landau bands (and in particular the same eigenfunctions), but with eigenvalues that generally differ from the canonical harmonic oscillator values. 
As a result, in the high magnetic field limit, the problem can be projected onto a single Landau level; but which one?
This depends on how the nontrivial dispersion rearranges the original Landau level energies.

To get more quantitative, we consider a zero-field Hamiltonian with an arbitrary, but isotropic kinetic energy term in two dimensions:
\begin{equation}
H_0=E_0 f \left( \frac{|\mb p|^2}{2mE_0} \right)\;, 
\label{eq:rings_H0}
\end{equation}
where $m$ is the free electron mass, $\mb p$ is the canonical momentum, and $E_0$ is an arbitrary energy scale that makes the argument of $f$ dimensionless. 
For non-relativistic free fermions, one has $f(x)=x$ (and the energy scale $E_0$ cancels out). 
Then, the eigenstates $\{ |\phi_{N,m} \rangle \}$ have eigenvalues 
 \begin{equation}
\epsilon_N^{\rm quadratic} = (N + 1/2) \hbar \omega_c = (N+ 1/2)\beta E_0 \;,
\label{eq:EVLL}
\end{equation}
where $N$ is the Landau level index, $m$ labels the intra-LL degeneracy, $\omega_c = eB/m$ is the cyclotron frequency, and $\beta$ is the dimensionless ratio $\hbar\omega_c / E_0$.
Since the Hamiltonian $H_0$ in \eq{rings_H0} is itself a function of $|\mb p|^2 /2m$, the $\{ |\phi_{N,m} \rangle \}$ are eigenstates for any $f(x)$:
\begin{equation}
H_0 | \phi_N \rangle = \epsilon_N |\phi_N\rangle,
\quad
\epsilon_N=E_0 f\left( (N+1/2)\beta \right) \;.
\label{energy}
\end{equation}
The simplest non-trivial example for $f(x)$ is
\begin{equation}
f(x) =   -C x +4x^2  \;, 
\end{equation}
which gives
\begin{equation}
H_0(C) = -C\frac{|\mb p |^2}{2m}+\frac{1}{E_0}\left(\frac{| \mb p |^2}{m} \right)^2 \, .
\label{eq:f1eqn}
\end{equation}
This kinetic energy yields a Fermi sea which is an annulus if $C>0$ and $\epsilon_F<0$, and a circle otherwise. 
The annular case can be seen as a simplified model for a realistic energy dispersion that has recently been observed~\cite{Jo2017B} for holes in GaAs quantum wells.
It is straightforward to find values of the parameter $C$ and Fermi energy $\epsilon_F$ where the system has an annular Fermi sea at zero field and the $N=0$ Landau level is the one with lowest energy, giving a CFL with a circular Fermi sea.
Breaking up the zero-field Fermi sea into multiple annular pieces~\cite{Ippoliti2017B} also yields a single, circular Fermi sea of CFs. 
In short, the CFL Fermi sea is completely insensitive to such isotropic distortions of the electron spectrum.

For certain choices of $f(x)$ it is also possible to have a different Landau level have minimum energy.
Then, one could obtain different phases, like the Moore-Read phase~\cite{Moore1991} (arising if $N=1$ has lowest energy) or symmetry-broken phases such as stripes or bubbles~\cite{Moessner1996} (if $N\geq 2$ has lowest energy).
This opens the exciting possibility of studying symmetry-broken phases in the extreme high-field limit, a regime opposite to the one in which they are normally observed.
In the simple example of \eq{f1eqn}, we find that these phases are predicted to occur at a realistic carrier density $\sim 10^{11} {\rm cm}^{-2}$ for band parameters $C \sim 3$ and $E_0 \sim 3 {\rm meV}$.
The recently observed annular Fermi sea for holes in GaAs quantum wells~\cite{Jo2017B} would most likely not give rise to these phases, due to the fact that the annular Fermi pocket is included in a larger Fermi sea ({\it i.e.} there is no gap between the band of interest and other bands).
On the other hand, band structures that are reasonably well approximated by \eq{f1eqn} are expected to occur quite generically in band-inverted semiconductors, such as HgTe, when a gap is opened {\it e.g.} by application of strain~\cite{Moon2006}. 


\section{Fermi pockets}
\label{sec:pockets}
\begin{figure}
\centering
\includegraphics[width=0.85\textwidth]{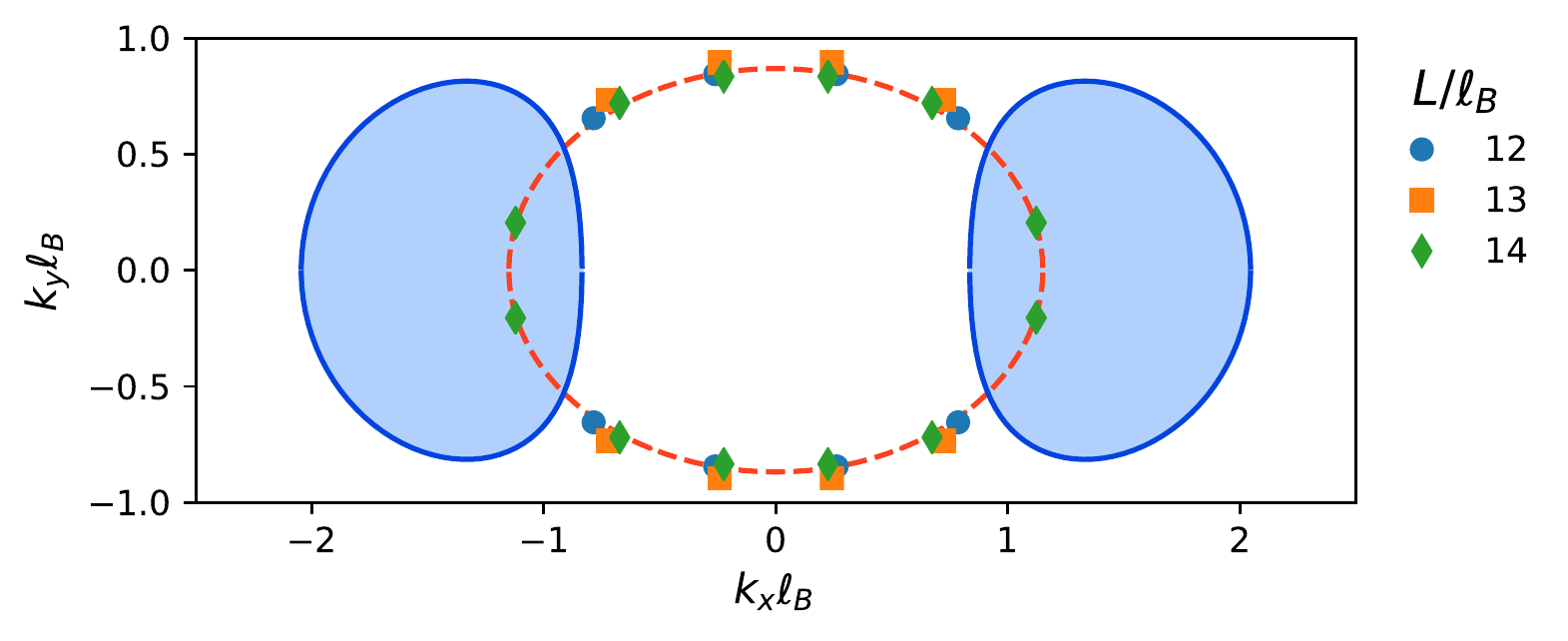}
\caption{Fermi contour of the zero-field electrons (solid blue lines) and $\nu = 1/2$ CFs (dashed red line). The latter is a fit to singularities in $S(q)$ (datapoints) obtained with iDMRG at the sizes indicated in the legend.
The Hamitonian is \eq{H_anis} with $\alpha=4$ and the Fermi energy $\epsilon_F=-0.49E_0$.
For these parameters, the inverse magnetic length at filling $\nu = 1/2$ is $\ell_B^{-1} = 0.64 k_0$,
so the two Fermi pockets are separated by $\approx 2\ell_B^{-1}$.
Nonetheless they give a connected Fermi sea, approximately elliptical, with anisotropy $\alpha_{CF} \simeq 0.76$.
}
\label{fig:beans}
\end{figure}

The previous Section shows that one can have an electron Fermi sea made of multiple disconnected pieces at zero field, and yet only a single CF Fermi sea at high field. 
We can observe a similar phenomenon for Fermi seas consisting of disconnected ``pockets'' without rotational symmetry, such as those shown in Fig.~\ref{fig:beans}. 
These Fermi contours are generated from the following zero-field Hamiltonian:
\begin{equation}
H_0(\alpha) = -\left( \frac{\alpha p_x^2}{2m} +\frac{p_y^2}{2m\alpha}\right) 
+\frac{1}{E_0}\left(\frac{p_x^2+p_y^2}{m} \right)^2\;.
\label{eq:H_anis}
\end{equation}
This is the same as \eq{f1eqn} with $C=1$, except that we have included an anisotropy parameter $\alpha$ in the quadratic part which explicitly breaks rotational symmetry.

Fermi contours with shapes similar to those in \fig{beans} have been observed in GaAs systems in a parallel field~\cite{Kamburov2014}, as well as in the surface states of Sn$_{1-x}$Pb${_x}$Se~\cite{Dziawa2012} and bismuth~\cite{Ohtsubo2012}~\cite{Feldman2016} (though in the latter case the valleys are elongated radially, rather than tangentially).
In such systems it is tempting to assume that multiple zero-field Fermi pockets imply that the system can be treated as having a single Fermi pocket and an additional ``valley pseudospin'' at high field.
However, we will show that this assumption is not always correct.

To prove this point, for the dispersion \eq{H_anis}, we choose the Fermi energy $\epsilon_F$ such that there are two disconnected zero-field pockets (as in Fig.~\ref{fig:beans}). 
Since we work at $\nu = 1/2 $, the Fermi energy also sets the cyclotron energy.
We solve the single-particle Schrodinger equation using \eq{H_anis} in the strong magnetic field determined by the $\epsilon_F$ and $\nu$, and find the two lowest-energy Landau bands.
These are even and odd under inversion, respectively, and in particular both of them have equal weight in the two valleys.
When their energies become nearly degenerate (i.e., when the splitting is sufficiently smaller than the interaction strength), the interactions can pick a linear superposition the Landau band which spontaneously breaks the inversion symmetry and populates only one (randomly chosen) Fermi pocket~\cite{Sodemann2017}. 
It would be a mistake, however, to assume that this hybridization always happens when the zero-field Fermi surface has multiple pockets. 
If the splitting between the two lowest-lying single-particle solutions is much larger than the strength of interactions, $\sim e^2/\ell_B \varepsilon$, then LL mixing effects are modest, and the quantum Hall problem takes place in one generalized LL that has equal weight in both valleys. 

We perform iDMRG calculations on the system described by \eq{H_anis} placed on an infinite cylinder with circumference $L = 13 \ell_B$. 
We retain the lowest two generalized LLs, project the Coulomb interaction accordingly, and introduce the single-particle energy splitting.
We set the total filling to be $\nu = 1/2$.
We obtain the many-body ground state and evaluate the difference in single-particle orbital occupations, $\delta n \equiv \langle \hat n_{{0}} \rangle - \langle \hat n_{{1}} \rangle$.
We find that that the system does \emph{not} valley-polarize until the interaction strength $e^2 / \ell_B $ is several times larger than the inter-Landau-level gap $\delta E$.
Nearly all electrons populate the inversion-even $N=0$ Landau level up to the value of $\alpha$ used in \fig{beans}, where the Fermi pockets are already clearly separated.
Even beyond that value, valley polarization is only possible in the immediate vicinity of isolated level crossings, or at very large values $\alpha \gtrsim 12$.

By studying the structure factor $S(\mb q)$ of states with $\delta n \simeq 1/2$, 
at sufficiently small $\alpha$, we are able to map the Fermi contour of CFs with the method discussed previously.
This shows that the composite Fermi contour goes from being a circle at the isotropic point (in agreement with the analysis of \sect{rings}) to an ellipse which becomes more elongated as the parameter $\alpha$ is increased. 
In \fig{beans} we show the results for $\alpha = 4$, where the CF Fermi contour is approximately elliptical, with a ratio of Fermi vectors $\alpha_{CF} \simeq 0.76$.
Thus, we again find a lack of a simple one-to-one correspondence between the Fermi contours of zero-field electrons and CFs:
while the electron Fermi sea is made of two well-separated pockets like in \fig{beans}, the CFL has a Fermi sea consisting of a \emph{single} connected component, close to an elliptical shape, and smoothly connected to the circular one at $\alpha=1$.

When $\alpha$ is tuned to extremely large values, we effectively have a bilayer system at total filling $\nu=1/2$, whose phase diagram may be quite complicated (and generally dependent on details of the model).
In particular, gapped Halperin 331 and 113 states~\cite{Halperin1983} would compete against the valley-polarized CFL.
The detailed study of this problem goes beyond the scope of this work.

\section{Higher-order rotational symmetry}
\label{sec:nfold}

In this Section we consider the effects of band distortions with discrete, $N$-fold rotational symmetry ($C_N$)~\cite{Ippoliti2017C}, which generalize the case of band mass anisotropy ($N=2$) studied in \sect{bandmass}.
There are generally two ways to introduce anisotropy in a quantum Hall system: 
either via the one-electron dispersion, or via the electron-electron interaction.
In the case of elliptical ($N=2$) anisotropy, the two approaches are equivalent, since a linear coordinate transformation can move the anisotropy between the band mass tensor $m_{ab}$ and the dielectric tensor $\varepsilon_{ab}$. 
This is not the case for generic anisotropy, and thus band anisotropy and interaction anisotropy are not equivalent.
Here we concentrate on the former. 
We consider a simple set of dispersions, one for each $C_N$ symmetry group with even $N$.
Completely general distortions such as those studied in experiments can then be thought of as arising from a sum of terms with different $C_N$ symmetries,
so the response of the CFL can be approximately broken down into its response to each individual ``harmonic''.
 
We quantify the anisotropy by $\alpha_F$, defined again as the ratio of the longest and shortest Fermi wavevectors for the system at zero magnetic field, directly generalizing the definition for band mass anisotropy.
The corresponding CF quantity, $\alpha_{CF}$, is defined analogously.
We study a class of model dispersions $\mathcal{E}(\mb k)$ that satisfy the following requirements:
\begin{enumerate}
\item $C_N$ symmetry.
\item Homogeneity: $\mathcal{E}(\mb k) = k^a \mathcal{E}(\hat{k} )$ (this guarantees the shape of the Fermi contour is independent of $\epsilon_F$).
\item Polynomial dependence on the components $k_x$, $k_y$ (this guarantees $\mc E$ is well-defined as an operator once the magnetic field is turned on, $\mb k \mapsto (\mb p -e \mb A)/\hbar$).
\item Positivity (this guarantees $\mc E$ has a finite-energy ground state). 
\end{enumerate}
Functions satisfying these requirements can easily be constructed by taking the exponent $a$ to be $N$:
\begin{equation}
\mathcal{E}_\lambda (\mb k) \equiv E_0  (\ell_0 k)^N (1+\lambda \cos(N\theta) ) \;,
\label{eq:disp}
\end{equation} 
where $(k,\theta)$ are polar coordinates in two-dimensional momentum space and $E_0$, $\ell_0$ are arbitrary units of energy and length.
This clearly has the desired $C_N$ symmetry.
Moreover, by observing that $k^N \cos(N\theta) = \text{Re} (k e^{i \theta})^N = \text{Re}(k_x + i k_y)^N$,
we see that this is a homogeneous polynomial of degree $N$ in $k_x$, $k_y$.
To ensure it is bounded from below we must choose $-1<\lambda < 1$.
The zero field electron Fermi contour for the dispersion in \eq{disp} is given by
\begin{equation}
k_F(\theta) = \frac{A}{\ell_0} \left( \frac{1}{1+\lambda \cos(N\theta)} \right)^{1/N}
\label{eq:kF_theta}
\end{equation}
where the dimensionless prefactor is $A= (\epsilon_F/E_0)^{1/N}$.
The dispersions $\mathcal{E}_{-\lambda}$ and $\mathcal{E}_{\lambda}$ are related by a $\pi/N$ rotation, so we can assume $\lambda \geq 0 $ without loss of generality.
The anisotropy parameter $\alpha_F$ can be expressed in terms of $\lambda$:
\begin{equation}
\alpha_F = \frac{k_F(\pi/N)}{k_F(0)} =  \left( \frac{1+\lambda}{1-\lambda} \right)^{1/N} \;. 
\label{eq:alpha_lambda}
\end{equation}
While the choice of dispersion \eq{disp} is guided by simplicity, we expect it to capture the main features of realistic distortions with the given symmetry. 
We thus contend that our results will be generic and not particular to this special class of anisotropic dispersions.

We numerically calculate $\alpha_{CF}$ as a function of $\alpha_F$ for $N = 4$ and 6 with the method from \sect{method}.
We do so by mapping the Fermi contour $k_{CF}(\theta)$ of composite fermions as a function of the angle $\theta$. 
We aggregate singularities from different system sizes (three to five distinct values of the cylinder circumference ranging between $13\ell_B$ and $21\ell_B$) at the same $N$ and $\alpha_F$.
The resulting data should satisfy \eq{kF_theta}, but with $\lambda$ replaced by a modified $\lambda_{CF}$, from which $\alpha_{CF}$ can be derived using \eq{alpha_lambda}. 

\begin{figure}
\centering
\includegraphics[width = \textwidth]{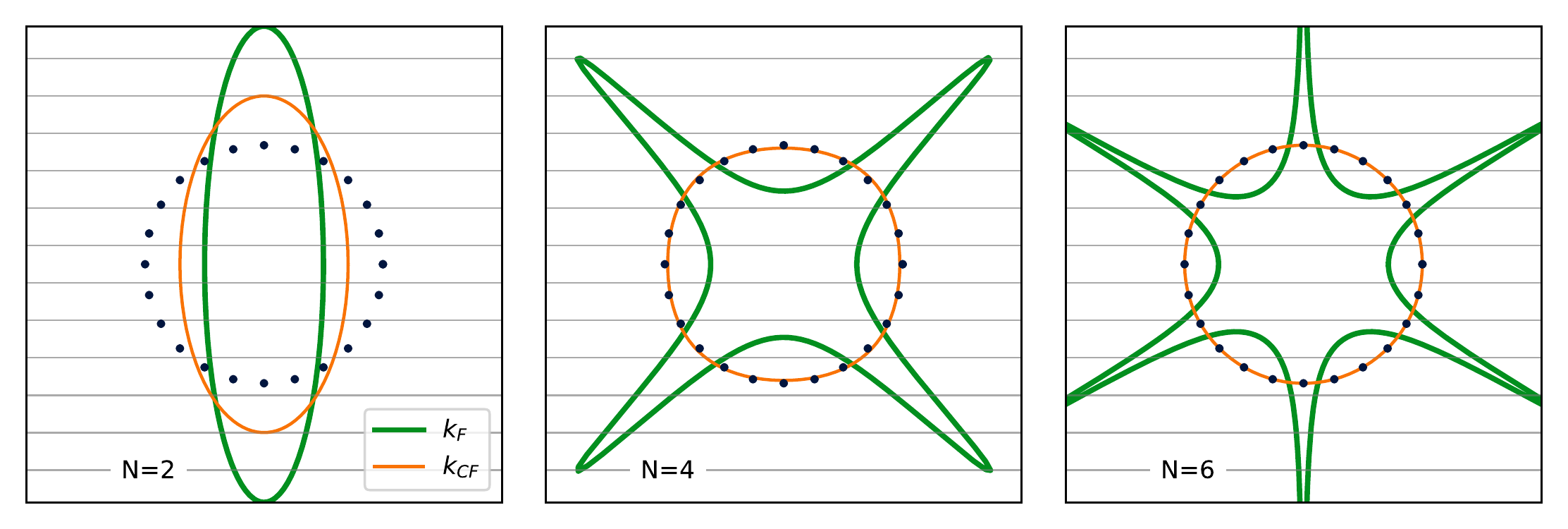}
\caption{
Fermi contours for the electrons (thick green) and the CFs (thin orange), for different values of $N$,
for the same value of zero-field Fermi contour anisotropy, $\alpha_F = 4$.
The dark blue dots correspond to a circular (isotropic) Fermi contour for reference. 
The gray lines indicate the accessible momenta on a cylinder of circumference $L_y=20 \ell_B$. 
}
\label{fig:surfaces}
\end{figure}

We find that the dependence of $\alpha_{CF}$ on $\alpha_F$ gets much weaker as $N$ is increased.
While the results for $N=2$ in \sect{bandmass} suggested $\alpha_{CF} \approx \alpha_F^{0.5}$, in this case we consistently find $\alpha_{CF}$ to be very close to 1 for $N = 4$ and 6.
In particular, the data for $N = 4$ deviates from a power law, 
possibly saturating at $\alpha_{CF} \simeq 1.05$.
For $N = 6$ the data is entirely consistent with $\alpha_{CF} = 1$ within numerical uncertainty.
The results are summarized in \fig{surfaces}.
The CF Fermi contour (thin orange line) for $N=6$ is not distinguishable from a circle (dark blue dots), whereas for $N=4$ there is a small but systematic difference, visible upon close examination.
This illustrates the drastic reduction in the magnitude of the effect when going from $N = 2$ to $N = 4$ and then $N=6$.

\section{Discussion \label{sec:nfold_conclusions}}

We have investigated the connection between the Fermi contours of zero-field carriers (electrons or holes) and their high-field CF counterparts. 
We divided our analysis into four cases of interest: 
band mass anisotropy, which is of direct experimental interest;
isotropic zero-field dispersions, potentially giving rise to a Fermi sea consisting of one or more annuli;
anisotropic dispersions giving rise to multiple disconnected pockets;
and anisotropic dispersions with higher-order discrete rotational symmetry ($C_N$ with $N>2$).

In general, we find the effects on the CFL to be quite subtle.
An exact analytical argument allows us to prove that the CFL is completely insensitive to the rotationally-symmetric distortion, so that its Fermi contour remains a circle, regardless of whether the system at zero field develops a disconnected Fermi sea.
A similar conclusion extends to the many-valley case, where the CFL either has a single connected Fermi sea, or may transition into different phases, depending on the separation of the pockets and the magnetic field.
In the case of higher-order rotational symmetry, while we do observe transference of the $C_4$-symmetric anisotropy to the CFL, the magnitude of the effect is very small.
For $N\geq 6$ the effect is smaller than our numerical accuracy, and thus most likely not measurable in experiment.

The model bands we have used to study the effects of $C_N$ symmetry were designed {\it ad hoc} and as such may be unrealistic. 
In particular, the absence of $k^2$ terms for $N>2$ is non-generic. 
However, restoring such isotropic terms would make the problem \emph{less} anisotropic overall.
Thus our model bands capture the essential features of distortions with the given symmetry, and a bound on their effect directly implies a bound on the effect of more realistic dispersions.
Our results show that the quadrupolar ($N=2$) distortion, corresponding to band mass anisotropy, has by far the strongest effect, with the effect of higher angular numbers decaying rapidly.
This indicates that finer structures in the zero-field problem are more efficiently washed out at high field.
This is a consequence of Landau-level projection (quantifiable by decomposing the projected Hamiltonian into anisotropic pseudopotentials~\cite{Ippoliti2017C}~\cite{BoYang2017A}), and is thus not exclusive to the CFL state.
Indeed, similar behavior is observed (through a different diagnostic) in incompressible states~\cite{Krishna2019}.

We conclude by mentioning recent work~\cite{Leaw2019} that finds a \emph{universal} Fermi contour anisotropy renormalization in two-dimensional, Coulomb-interacting Dirac systems.
The result is seemingly consistent with the behavior of the CFL in the presence of band mass ansiotropy, $\alpha_{CF} \approx \sqrt{\alpha_F}$, as seen both in experiment and in our simulations. 
The connection between these results is unclear, as the CFs are \emph{not} expected to interact via the Coulomb interaction, and their Dirac nature is not conclusively established.
Either way, explaining properties of the CFL through general Fermi liquid physics is a fascinating direction for future work, even beyond the geometric distortions studied here.
In particular, this may lead to a better understanding the excitation spectrum of this peculiar quantum fluid.

\begin{acknowledgements}
We acknowledge support from DOE BES grant DE-SC0002140.
RNB acknowledges the hospitality of the Aspen Center for Physics during the writing of this manuscript.
The work presented was done in collaboration with Scott Geraedts.
The infinite DMRG libraries used in this work were created by Michael Zaletel, Roger Mong and the TenPy collaboration.
We thank Mansour Shayegan for useful discussions.
\end{acknowledgements}

\bibliographystyle{spphys}       
\bibliography{qfs}   

\end{document}